\documentclass{article}

\usepackage{arxiv}
\usepackage{mathtools}
\usepackage[utf8]{inputenc} 
\usepackage[T1]{fontenc}    
\usepackage{hyperref}       
\usepackage{url} 
\usepackage{booktabs}       
\usepackage{amsfonts}       
\usepackage{nicefrac}       
\usepackage{microtype}      
\usepackage{lipsum}		
\usepackage{graphicx}
\usepackage{natbib}
\usepackage{doi}
\usepackage{amsmath}
\usepackage{graphicx}
\usepackage{caption}
\usepackage{subcaption}
\usepackage{multirow}
\usepackage{float}
\usepackage{listings}

\usepackage{authblk}
\title{JetFlow: generating jets with conditioned and mass constrained normalising flows }

\author[1,*]{Benno Käch}

\author[1]{Dirk Krücker}

\author[1]{Isabell-A Melzer-Pellmann}

\author[1,2,3]{Moritz Scham}

\author[1,2]{Simon Schnake}

\author[1,2]{Alexi Verney-Provatas}

\affil[1]{Deutsches Elektronen-Synchrotron DESY\\
	 Germany}
\affil[2]{RWTH Aachen University, III. Physikalisches Institut A, Aachen, Germany}
\affil[3]{Institute for Advanced Simulation - Jülich Supercomputing Centre, Juelich, Germany}
\affil[*]{Corresponding Author: benno.kaech@desy.de}

\renewcommand{\shorttitle}{\textit{arXiv} Template}

\hypersetup{
pdftitle={JetFlow},
pdfauthor={Benno Käch, Dirk Krücker, Isabell Melzer-Pellmann},
pdfkeywords={jetnet, normalising flows},
}
\setcitestyle{numbers,square}
\begin{document}
\maketitle

\begin{abstract}
 Fast data generation based on Machine Learning has become a major research topic in particle physics. This is mainly because the Monte Carlo simulation approach is computationally challenging for future colliders, which will have a significantly higher luminosity. The generation of collider data is similar to point cloud generation with complex correlations between the points.
	In this study, the generation of jets with up to 30 constituents with Normalising Flows using Rational Quadratic Spline coupling layers is investigated. Without conditioning on the jet mass, our Normalising Flows are unable to model all correlations in data correctly, which is evident when comparing the invariant jet mass distributions between ground truth and generated data. Using the invariant mass as a condition for the coupling transformation enhances the performance on all tracked metrics. In addition, we demonstrate how to sample the original mass distribution by interpolating the empirical cumulative distribution function. Similarly, the variable number of constituents is taken care of by introducing an additional condition on the number of constituents in the jet.
	Furthermore, we study the usefulness of including an additional mass constraint in the loss term. On the \texttt{JetNet} dataset, our model shows state-of-the-art performance combined with fast and stable training.

Significance: The contribution demonstrates that Normalising Flows with Rational Quadratic Splines can model high-dimensional data efficiently (i.e. fast training and state-of-the-art performance) when global features (mass) are used for conditioning the transformation.

Reference: The study uses a the public \texttt{JetNet} dataset: \url{https://zenodo.org/record/4834876} and arXiv:2106.11535 
\end{abstract}


\section{Introduction}
	High-Energy Physics (HEP) has benefited from the advances in Machine Learning (ML), since the analysis of HEP data is a high-dimensional multivariate problem. ML has been applied in multiple ways ranging from classification to regression tasks~\cite{mlinhep}. In HEP, detailed simulations of the physical processes are common, which almost perfectly describe the details of the experimental measurement. These Monte Carlo simulations (MC) provide labeled data and are needed in large numbers to cover the extreme areas of the physical phase space. The simulations for the CMS detector at the Large Hadron Collider, for example, require about $50 \%$~\cite{mccms} of the current CMS computing budget. An even larger number of simulations will be needed for the upcoming high-luminosity phase of the LHC~\cite{hllhc}. Therefore, generative modelling with Deep Learning (DL) sparked great interest in the HEP community.

In this study, the generation of jets is investigated using different Normalising Flow (NF)~\cite{tabak,rezende,kobyzev_2021,papama19} architectures.  The use of NFs is motivated by their stable convergence thanks to Maximum Likelihood training, however NFs often struggle to model all correlations correctly. One powerful summary statistic in this study is the invariant mass of the jet, which depends on the features of all constituents forming the jet. Thus the distribution for the mass calculated from generated and MC simulated jets are required to be compatible.
 This paper first provides a summary of related work in the field of generative models for jet generation in Section~\ref{sec:related}, which also introduces the dataset on which this research was conducted. 
This is followed by a description of the architecture of the models proposed in Section~\ref{sec:Architecture}. In Section~\ref{sec:results} the generated data is presented and visually compared to holdout data for different architectures. The results are then compared to state-of-the-art results from the field and are briefly discussed in Section~\ref{sec:discussion}. Finally, a conclusion is given in Section~\ref{sec:conclusion}, which highlights the important aspects of this approach and plans for future work.
\section{Related Work}
\label{sec:related}
The ML-aided generation of physical processes is studied in~\cite{naturegen}, where Generative Adversarial Models (GANs) and Variational Auto-Encoders are compared on problems with a wide range of dimensionalities,  from a toy model with 10 dimensions to 26-dimensional physical data from the Z decay. In their work, the authors show that with their model, they are able to model the data distribution to an acceptable degree. However, this problem becomes increasingly more difficult in a higher number of dimensions, a problem that was also observed during this study.
Here, we consider the generation of jets with up to 30 particles - leading to a total number of 90 dimensions which need to be generated. Given that the generation of images of faces is possible~\cite{glow}, which needs orders more dimensions, one could expect that this would be an easy task. However, it proves difficult, as there are sensitive summary statistics available in our case. In Ref.~\cite{jetnetlib} a dataset of particle clouds, each containing up to 30 particles and each yielding 3 features, is provided allowing benchmarking of different generative models. The library's creators also published a paper with a message-passing GAN~\cite{mpgan}, which outperforms other models by orders of magnitude on nearly every metric. This study contributes to a comparison of different NF approaches on this problem, whereas in the previous publication only GANs are considered. 
\subsection{Discrete Normalising Flows}
Discrete Normalising Flows are  a relatively new subject in HEP, an conclusive overview is given in Refs.~\cite{kobyzev_2021} and \cite{papama19}.
Invertible and differentiable transformations $\boldsymbol{f}$ are sought, which transform the training data distribution $X$ to a Normal Gaussian distribution $Z$. The change of variables formula for probability distribution given in Equation~\ref{eq:change}, motivates the objective of the training, as the goal is to maximize the likelihood of the transformed data under a Normal distribution. 
\begin{equation}
    p_X(\boldsymbol{x})=p_Z(\boldsymbol{f}(\boldsymbol{x}))\left |\textrm{det }D\boldsymbol{f}(\boldsymbol{x})) \right|
    \label{eq:change}
\end{equation}
This is the normalising direction of the NF, which is used to train the model. Note that maximizing the likelihood in the input space is not possible, as we cannot evaluate the value of the probability of the data distribution $p_X(\boldsymbol{x})$. The concatenation of invertible transformations is itself also invertible, thus multiple transformations are chained in sequence to obtain an overall more expressive transformation. The loss function of the model can be constructed from Equation~\ref{eq:change} by minimizing the negative log-likelihood of $M$ samples of training data in the input space (for simplicity only one transformation is used, but for multiple transformations, the individual terms can be summed to obtain the loss of the composite transformation):
\begin{equation}
    L_{nLL}(\boldsymbol{\theta})=-\sum_{i=1}^M \textrm{log } p_X(\boldsymbol x^{(i)}|\boldsymbol\theta)=-\sum_{i=1}^M \textrm{log } p_Z(\boldsymbol{f}(\boldsymbol{x}^{(i)}|\boldsymbol\theta))-\textrm{log }\left |\textrm{det }D\boldsymbol{f}(\boldsymbol{x|\boldsymbol\theta})
    \right|
    \label{eq:loss}
\end{equation}

If the model can transform the data to a Gaussian sufficiently well, sampling the data distribution is straightforward. First, a sample from the Normal distribution is drawn, and by applying the inverted transformations a sample from the data distribution is obtained. In addition to the constraint of only applying bijective transformations, it is also required that the determinant of the Jacobian is calculable efficiently. Generally, the calculation of the determinant of a $D\times D$ matrix needs $O(D^3)$ computations. For triangular matrices, this is significantly faster, as in this case the determinant is given by the product of its diagonal elements. Thus, NFs usually rely on transformation with a triangular Jacobian. This leads to two types of NF classes: models relying on autoregressive transformation as in~\cite{made,caloflow} and in contrast, models that use coupling layers~\cite{glow,realnvp}. For coupling layers the transformations are additionally applied element-wise and the transformations are constructed such that the Jacobian is an upper triangular matrix for which the determinant is just the product of its diagonal elements.

\subsection{Dataset}
In this study, the \texttt{JetNet}~\cite{mpgan} datasets are used. 
There are 3 different datasets available, each containing jets with an energy of about 1~TeV, with each jet containing up to 30 constituents. The difference in the datasets lies in the  jet-initiating parton. Datasets for top quark, light quark and gluon initiated jets are studied.
The jet constituents are considered to be massless and can therefore be described by their 3-momenta or equivalently by transverse momentum $p_T$, pseudorapidity $\eta$, and azimuthal angle $\phi$. In the \texttt{JetNet} dataset, these variables are given relative to the jet momentum: 
$\eta_i^\mathrm{rel} \coloneqq \eta_i^\mathrm{particle}\!-\eta^\mathrm{jet}$,
$\,\phi_i^\mathrm{rel} \coloneqq (\phi_i^\mathrm{particle}\!-\phi^\mathrm{jet})\bmod 2\pi$, and 
$\,p_{T,i}^\mathrm{rel} \coloneqq p_{T.i}^\mathrm{particle}\!/p_T^\mathrm{jet}$
, where $i$ runs over the particles in a jet.

The invariant mass $m_\mathrm{jet}$ of a jet is an essential high-level feature containing important physics information. It is a global variable that depends on the correlations between the single jet constituents and provides therefore an important metric for the performance of the generative model. For the relative quantities above, we can define the relative jet mass as $(m^\mathrm{rel})^2=(\sum_i E^\mathrm{~rel}_i)^2-(\sum_i \vec{p}^\mathrm{~rel}_i)^2 = m_\mathrm{jet}^2/p_\mathrm{T,jet}^2$.

\section{Architecture}
\label{sec:Architecture}
Although autoregressive models are strongly motivated by the decomposition of joint probability distributions into conditional ones. But, either the generating or the normalising direction of the NF is $D$ times slower, where $D$ is the dimension of the input space. In the scope of the study, this would mean that one direction is 90 times slower than the other. Ref.~\cite{caloflow2} proposed a solution to this issue, however as we introduce the mass constraint later, this solution is not viable anymore during training.
\subsection{Normalising Flows with coupling layers}
These NFs consists of multiple coupling layers, each coupling layer splits the input features into two (in this study equally sized) sets. The former are mapped with an identity transformation, while to  the latter a non-trivial parameterised element-wise transformation is applied. The parameters are the output of a neural network taking as input the former set. The splits of the two sets are chosen randomly for each layer at the beginning of the training and stay fixed from then on. An affine transformation is both simple and powerful, with the coupling layer having the following illustrative form:
\begin{equation}
    f(\boldsymbol x)=\begin{cases}
    \boldsymbol z^A=\boldsymbol x^A &\\ 
    \boldsymbol z^B=\boldsymbol x^B\odot s_{\boldsymbol\theta(\boldsymbol x^A)}+t_{\boldsymbol\theta(\boldsymbol x^A)} &
    \end{cases}
\end{equation}
This yields the following form for the determinant:
\begin{equation*}
    D\boldsymbol f=\begin{bmatrix}
        \mathbb{I} & 0 \\
        \frac{\partial \boldsymbol{z}_{D/2:D}}{\partial \boldsymbol{x}_{1:D/2}} & \text{diag}(s_{\boldsymbol\theta}(\boldsymbol x_{1:D/2}))
    \end{bmatrix}
    \Rightarrow\textrm{det}|D\boldsymbol f|=\prod^{D/2}_{k=1}s_{\boldsymbol\theta(\boldsymbol x_{1:D/2})_k} 
\end{equation*}
    
Here $D$ is the dimension of the input space and $s,t$ are arbitrary complex neural networks. This transformation is invertible as long $s\neq 0$, which is achieved by exponentiation of the output of the network.
In this study, more complex coupling layers were studied, namely rational quadratic monotonic splines~\cite{nsf}. The splines have $K$ bins and the bound of the splines is given by $[-B, B]$. In this study, the number of bins was scanned from $4-7$, while the bound $B$ was scanned from $3-10$. The rational quadratic splines require $3K+1$ parameters to define the transformation, as $K$ bins require $2K+2$ bin coordinates to fix the bin width and height. Further $K-1$ derivatives $\delta^{(k)}$ define the non-trivial part of the spline since outside the bound the spline is defined as the identity transformation, thus setting the derivatives $\delta^{(0)}=\delta^{(K)}=1$ for continuous bin edges. The \texttt{nflows}~\cite{nflows} library has these coupling layers and a general NF framework already implemented. The NNs consist of $l$ fully-connected feed-forward layers, using residual blocks relying on skip connections. These residual blocks allow a more stable training, leading to fewer spikes in the training loss. The activation function used during training is ReLU. The number of coupling layers was scanned from 5 to 30, and the number of residual blocks was scanned from 1 to 5. The model uses the Adam optimizer~\cite{adam}, the learning rate was scanned from 0.0005 to 0.001. It proved useful to use dropout during training, along with batch normalisation.

\subsubsection{Mass Conditioning}
Normalising flows are conditioned by adding variables to the input of the NNs  predicting the coupling transformations' parameters. These additional variables are not transformed. Conditioning enhances the expressivity of the transformation, as more information is given to the network predicting the parameters of the coupling layers. In the case of this study the invariant jet mass $m$ and optionally the number particles per jet $n$ are used for conditioning. In the following, the variables that are used to condition are given in brackets after the name of the model. It is important to note at this point that if we condition on a variable we also need to supply it when the NF is used in the generating direction.Thus the conditioning variables need to be generated independently as well. In the case of one conditioning variable, this is done by transforming the variable with its cumulative distribution function to a uniform distribution from which sampling is easy. This one-dimensional transformation is then interpolated with monotonic piece-wise cubic Hermite polynomials~\cite{pchip}. As it is monotonic and one-dimensional, it is easily invertible. Thus sampling with one condition is trivial. If two conditions are used, the two-dimensional space can either be modelled with a second normalising flow, or alternatively an autoregressive approach can be used: first the number of particles is sampled from its empirical probability mass function and for each value of number particles the one-dimensional approach is used independently. This autoregressive approach is especially viable here, because the number particles per jet is discrete and thus there are a finite number of conditional mass distributions. 

\subsubsection{Mass Constraint}
To further improve the mass modelling a mass constraint is introduced. Here, we add another loss term: the $L_2$ Norm between the mass a generated jet has been conditioned with and the value of the mass that is calculated from its generated features:
\begin{equation}
    L_\mathrm{MSE}=|m_\mathrm{cond}-m_\mathrm{gen}(\boldsymbol x_\mathrm{gen})|^2
\end{equation}
It is important to note that this means that the flow is used in both directions during one training iteration. 

\subsection{Variable Sized Particle Clouds}
As NFs need to be invertible, there are major constraints on the dimension of the generated data space, seemingly disallowing the generation a variable amount of particles. In this study, this was resolved by zero-padding point clouds with fewer than 30 particles. Noise in the order of $O(10^{-7})$ was added to the zero-padded particles as otherwise, it is an ill-posed problem in the context of normalising flows. For the case where the number particles is given as a condition, the particles higher than the condition are set to zero.

\section{Results}
\label{sec:results}
In this section results for 3 different architectures are presented. For each architecture, an equally sized random search was conducted. Results are presented for all 3 datasets. In terms of speed, the model needs  $9.77 \pm 0.04$ $\mu s$ for the sampling of a jet on an NVIDIA P100 when using a batch size of 50000. The training time for all models ranges from 1 to 2 hours on an NVIDIA P100.

\subsection{Evaluation Metrics}
A major problem in generative modelling is the absence of a general metric to evaluate the performance of the model. The Wasserstein distance would be a promising candidate to compare the generated distribution with the training distributions. Unfortunately it is not tractable in general in more than one dimension. Thus the Wasserstein distances $W_1$ between the inclusive distributions of the MC-simulated and the flow-generated jets are used for hyperparameter optimisation. These are inclusive in a sense that the average of the Wasserstein distance is calculated between the distributions of $(\eta^\mathrm{rel},\phi^\mathrm{rel},p_T^\mathrm{rel})$ of all particles, not the individual ones. However, only checking the distributions of the generated features is not enough, since the correlations between the individual particles must be represented. One feature describing a complex correlation in this study is the invariant mass of the jet. To quantify the agreement $W_1^M$, the Wasserstein distance between the NF generated mass distribution and the ground truth (i.e.\ MC simulated) mass distribution is used. Additionally, the Wasserstein distance $W_1^{EFP}$ is considered, which calculates the Wasserstein distance between energy flow polynomials for ground truth and generated jets. Energy flow polynomials~\cite{efp} form a basis for all useful jets observables, thus a good agreement on this basis is required.

Further metrics used are computer-vision inspired, as for coverage (COV) and minimum matching distance (MMD)\cite{covmmd}. Kansal et al. introduce a modified version of the Fréchet Inception Distance~\cite{FID} called Fréchet ParticleNet Distance (FPND)~\cite{jetnet}, which calculates the Wasserstein-2 distance between Gaussians fitted in the first fully connected layer of a ParticleNet classifier~\cite{particlenet} for ground truth and generated samples.

\subsection{Performance comparison}

Since the Wasserstein distances have significant statistical fluctuation, an average of over five samples is presented, together with its standard deviation. The distances were calculated the same way as in~\cite{mpgan}, i.e. the Wasserstein distances in batches of 10000 and the FPND between 50000 ground truth and generated jets. The results for the models with the lowest $W_1^M$ validation score of the investigated architectures are shown in Table~\ref{tab:results}. These are the NF without conditioning (Vanilla NF, short VNF), NF with conditioning (CNF), and NF with conditioning and mass constraint (CCNF). For architectures that rely on conditioning, the variables that were used for conditioning are given in brackets.

\begin{table}[h]
\centering
\begin{tabular}{@{}ll|lllllll@{}}
\toprule

Jet Class & Model & $W_1^M (\times 10^{-3})$ & $W_1^P
(\times 10^{-3})$ & $W_1^{EFP}(\times 10^{-5})$ &  FPND & COV $\uparrow$ &   MMD \\
\cline{1-8}
\multirow{7}{*}{Gluon} & 
MP-MP &$0.7 \pm 0.2$ &$0.9 \pm 0.3$ &$\mathbf{0.7 \pm 0.7}$ & $\mathbf{0.12}$ & $\mathbf{0.56}$ &$0.037$ \\&

MP\_LFC-MP &$0.69 \pm 0.07$ &$1.8 \pm 0.3$ & $0.9 \pm 0.2$ &$0.20$ &$0.54$ &$0.037$ \\&
VNF &$4.3 \pm 0.2$ &$2.2 \pm 0.5$ & $3 \pm 1$ &$1.82$ &$0.53$ & $\mathbf{0.035}$ \\&
 CNF\ (m) &$0.9 \pm 0.3$ & $\mathbf{0.5 \pm 0.2}$ & $0.8 \pm 0.6$ &$0.53$ & $\mathbf{0.56}$ &$0.036$ \\&
 CNF\ (m,n) &$0.8 \pm 0.3$ &$1.2 \pm 0.3$ & $0.8 \pm 0.8$ &$0.31$ &$0.55$ & $\mathbf{0.035}$ \\&
CCNF\ (m) & $\mathbf{0.6 \pm 0.2}$ &$0.6 \pm 0.2$ & $1.2 \pm 0.9$ &$0.90$ &$0.54$ &$0.036$ \\&
CCNF\ (m,n) &$0.7 \pm 0.4$ &$0.9 \pm 0.3$ & $1.1 \pm 0.6$ &$0.54$ & $\mathbf{0.56}$ &$0.036$ \\\cline{1-8}
\multirow{7}{*}{Light Quark} & 
MP-MP & $\mathbf{0.6 \pm 0.2}$ &$4.9 \pm 0.5$ &$\mathbf{0.7 \pm 0.4}$ &$0.35$ &$0.50$ &$0.026$ \\&

MP\_LFC-MP &$0.7 \pm 0.2$ &$2.6 \pm 0.4$ & $0.9 \pm 0.9$ & $\mathbf{0.08}$ &$0.52$ &$0.037$ \\&
VNF &$2.8 \pm 0.6$ &$2.2 \pm 0.4$ & $1.9 \pm 0.6$ &$1.30$ & $\mathbf{0.54}$ & $\mathbf{0.024}$ \\&
 CNF\ (m) &$0.9 \pm 0.3$ &$1.1 \pm 0.4$ &$\mathbf{0.7 \pm 0.3}$ &$0.49$ &$0.53$ & $\mathbf{0.024}$ \\&
 CNF\ (m,n) &$1.0 \pm 0.2$ &$4.2 \pm 0.6$ &$\mathbf{0.7 \pm 0.4}$ &$0.73$ &$0.53$ & $\mathbf{0.024}$ \\&
CCNF\ (m) &$0.7 \pm 0.2$ & $\mathbf{0.8 \pm 0.5}$ &$\mathbf{0.7 \pm 0.4}$ &$0.55$ &$0.50$ &$0.025$ \\&
CCNF\ (m,n) &$0.7 \pm 0.1$ &$4.5 \pm 0.9$ & $1.0 \pm 0.6$ &$1.22$ &$0.50$ &$0.025$ \\\cline{1-8}
\multirow{7}{*}{Top Quark} & 
MP-MP &$0.6 \pm 0.2$ &$2.3 \pm 0.3$ &$\mathbf{2 \pm 1}$ & $\mathbf{0.37}$ &$0.57$ &$0.071$ \\&

MP\_LFC-MP &$0.9 \pm 0.3$ &$2.2 \pm 0.7$ &$\mathbf{2 \pm 1}$ &$0.93$ &$0.56$ &$0.073$ \\&
VNF &$6.6 \pm 0.6$ &$2.2 \pm 0.5$ &$15 \pm 1$ &$7.76$ & $\mathbf{0.59}$ & $\mathbf{0.070}$ \\&
 CNF\ (m) &$1.8 \pm 0.5$ & $\mathbf{1.2 \pm 0.4}$ & $3 \pm 1$ &$2.62$ &$0.57$ & $\mathbf{0.070}$ \\&
 CNF\ (m,n) & $\mathbf{0.55 \pm 0.08}$ &$1.7 \pm 0.3$ & $3 \pm 1$ &$2.19$ &$0.57$ &$0.071$ \\&
CCNF\ (m) &$0.7 \pm 0.2$ & $12.2 \pm 0.5$ & $8 \pm 3$ &$8.52$ &$0.44$ &$0.076$ \\&
CCNF\ (m,n) &$1.1 \pm 0.5$ &$1.3 \pm 0.4$ & $4.9 \pm 0.7$ &$2.27$ &$0.56$ &$0.073$ \\
 \bottomrule
\end{tabular}
\vspace{20 pt}
\caption{Comparison between the best performing models from~\cite{mpgan}, with 5 different model from this study. No systematic differences can be seen except for the FPND metric and $W_1^\mathrm{EFP}$ on the top-quark data set.}
\label{tab:results}

\end{table}



\begin{figure}[h]
    \centering
    \includegraphics[width=\textwidth]{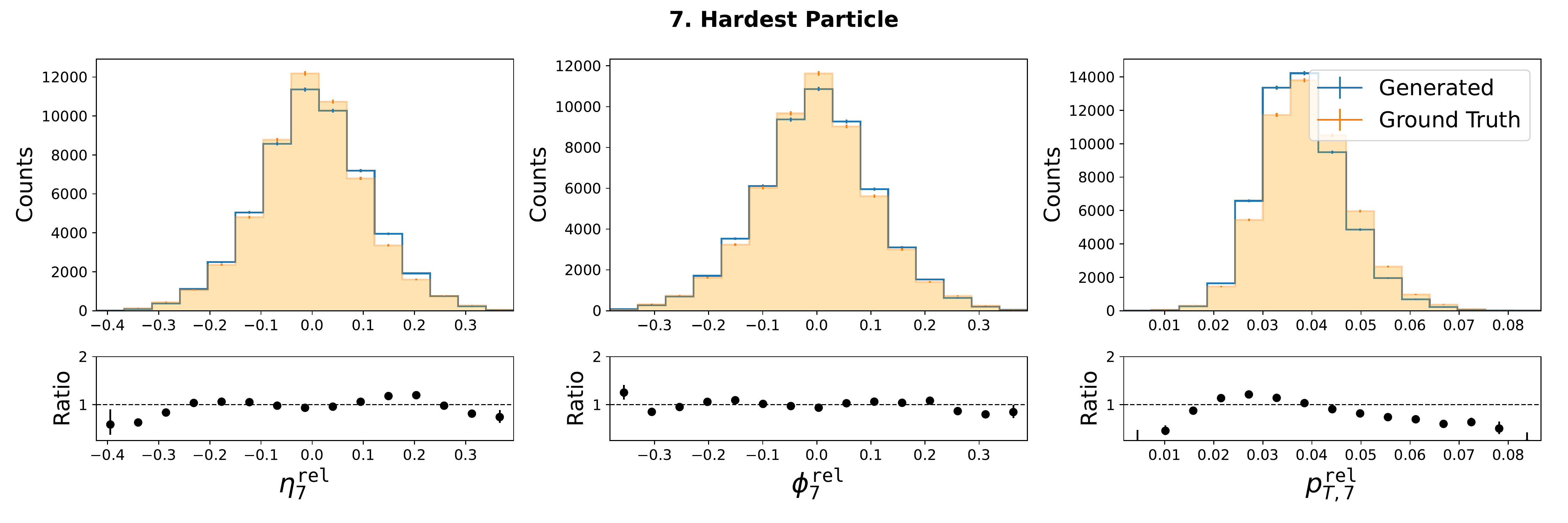}
    \caption{Marginal histograms of $(\eta^\mathrm{rel}, \phi^\mathrm{rel}, p_T^\mathrm{rel})$ for the $7^{th}$ hardest particle. Shown are examples for the top-quark dataset, which is trained with the CCNF architecture.}
    \label{fig:marginals}
\end{figure}

As an example, marginal distributions for the individual particle in the jet are shown for the 7th hardest particle in Fig.~\ref{fig:marginals} for the top-quark dataset, which is trained with the CCNF architecture. Distributions for the other jet constituents behave similar.

\begin{figure}[h]
    \centering
    \includegraphics[width=\textwidth]{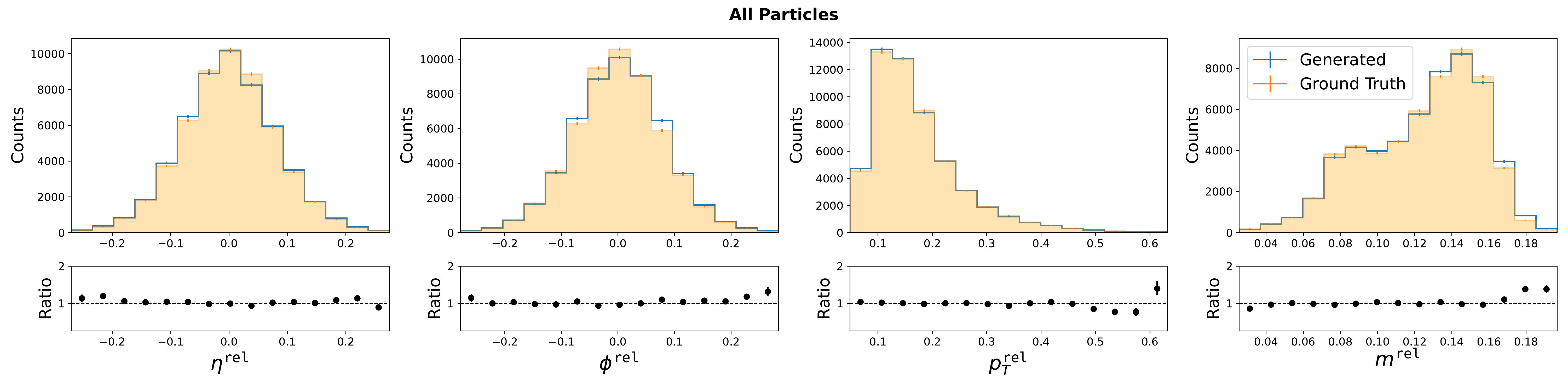}
    \caption{Inclusive histograms of $(\eta^{rel},\phi^{rel},p_T^{rel},m^{rel})$ for all particles. Shown are examples for the top-quark dataset which was trained with the CCNF architecture using 2 conditions.}
    \label{fig:inclusive}
\end{figure}

The inclusive distribution of the training features as well as the invariant mass distribution for the top-quark dataset are presented in Fig.~\ref{fig:inclusive}. These distributions are used for the calculation of $W_1^P$, which is done for the inclusive distribution over all particles in the jet. 

\begin{figure}[h]
    \centering

\begin{subfigure}{0.33\textwidth}
    \centering
    \includegraphics[width=\textwidth]{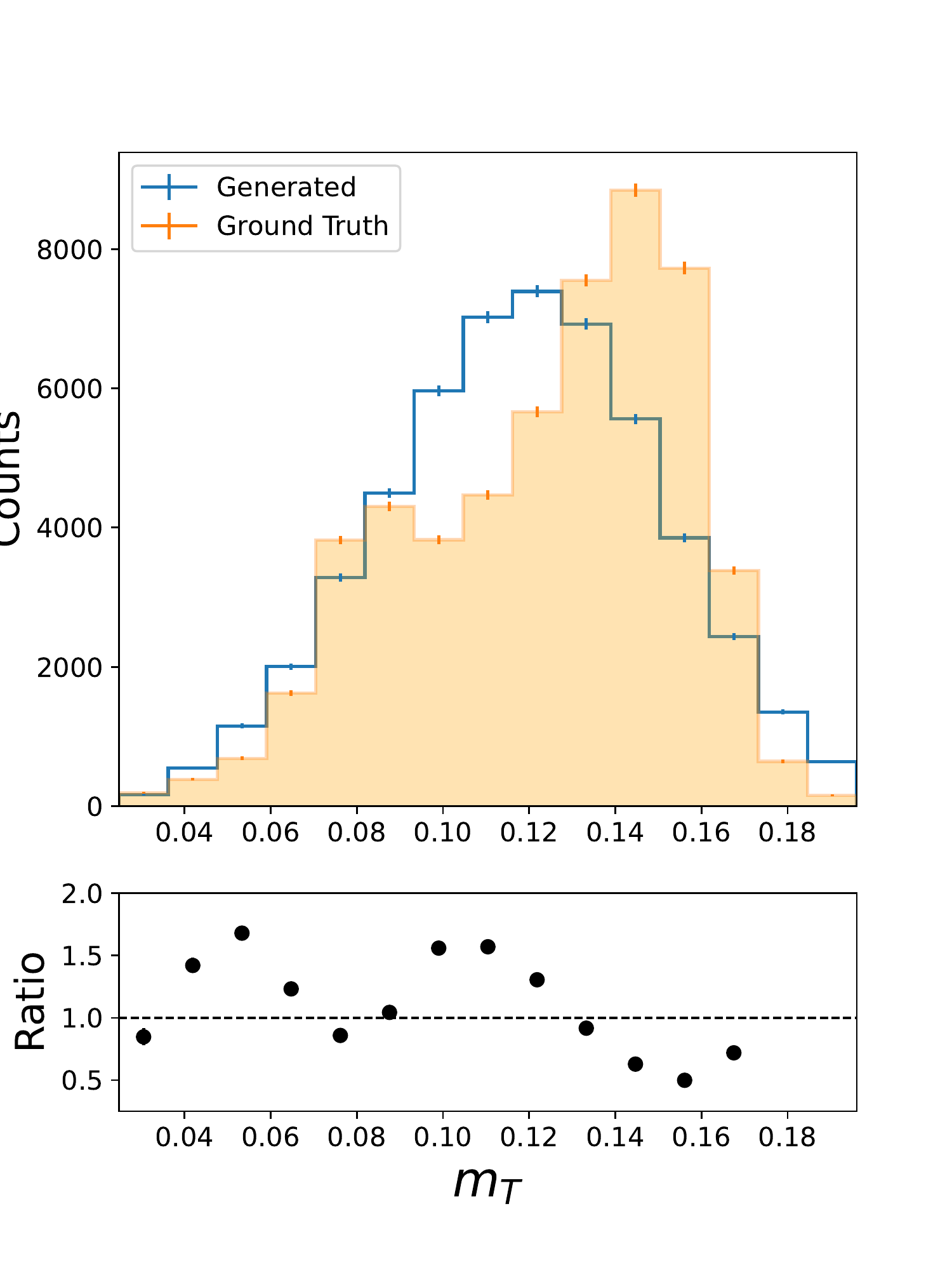}

\end{subfigure}
\begin{subfigure}{0.33\textwidth}
    \centering
    \includegraphics[width=\textwidth]{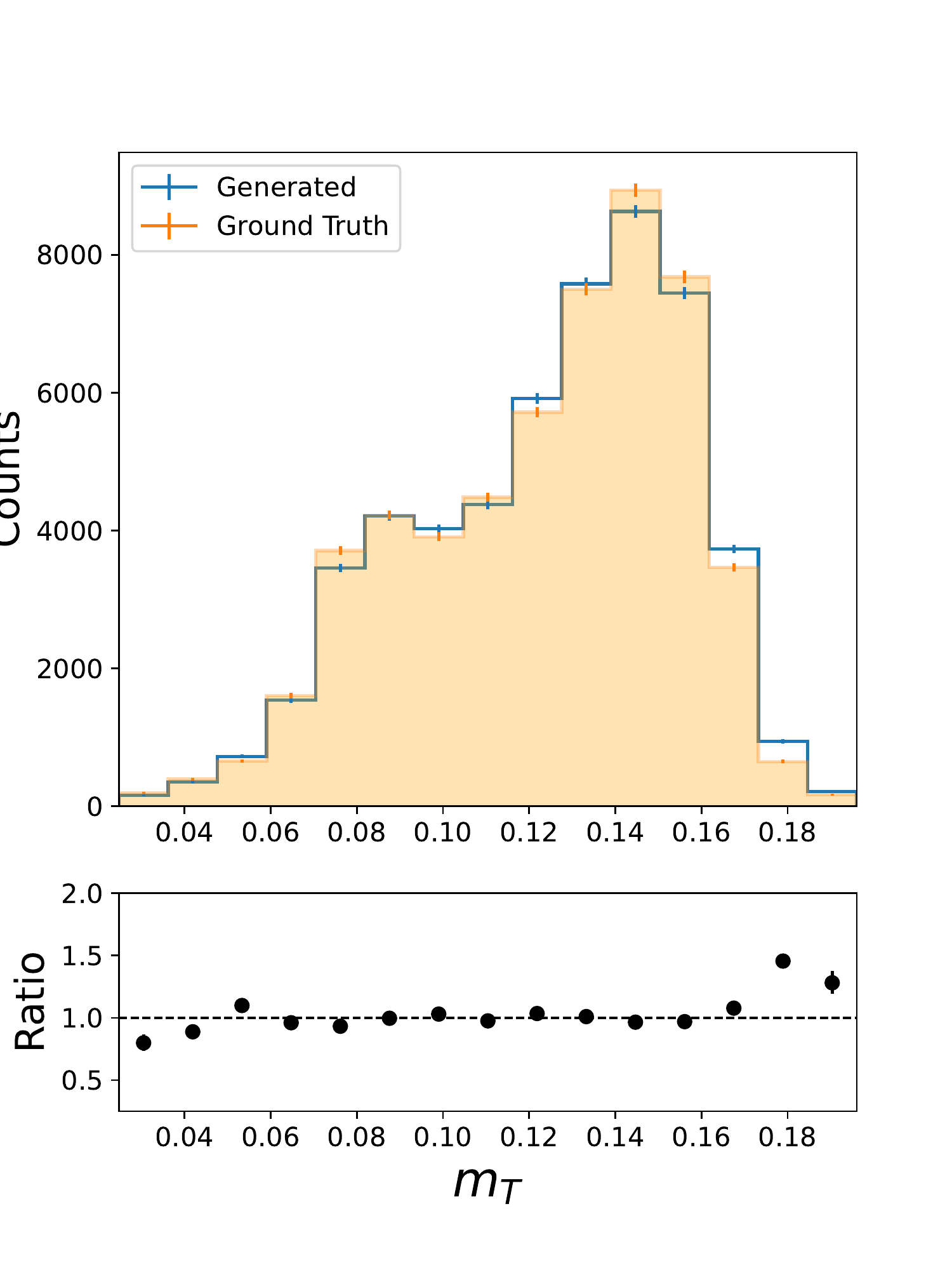}

\end{subfigure}
\begin{subfigure}{0.33\textwidth}
    \centering
    \includegraphics[width=\textwidth]{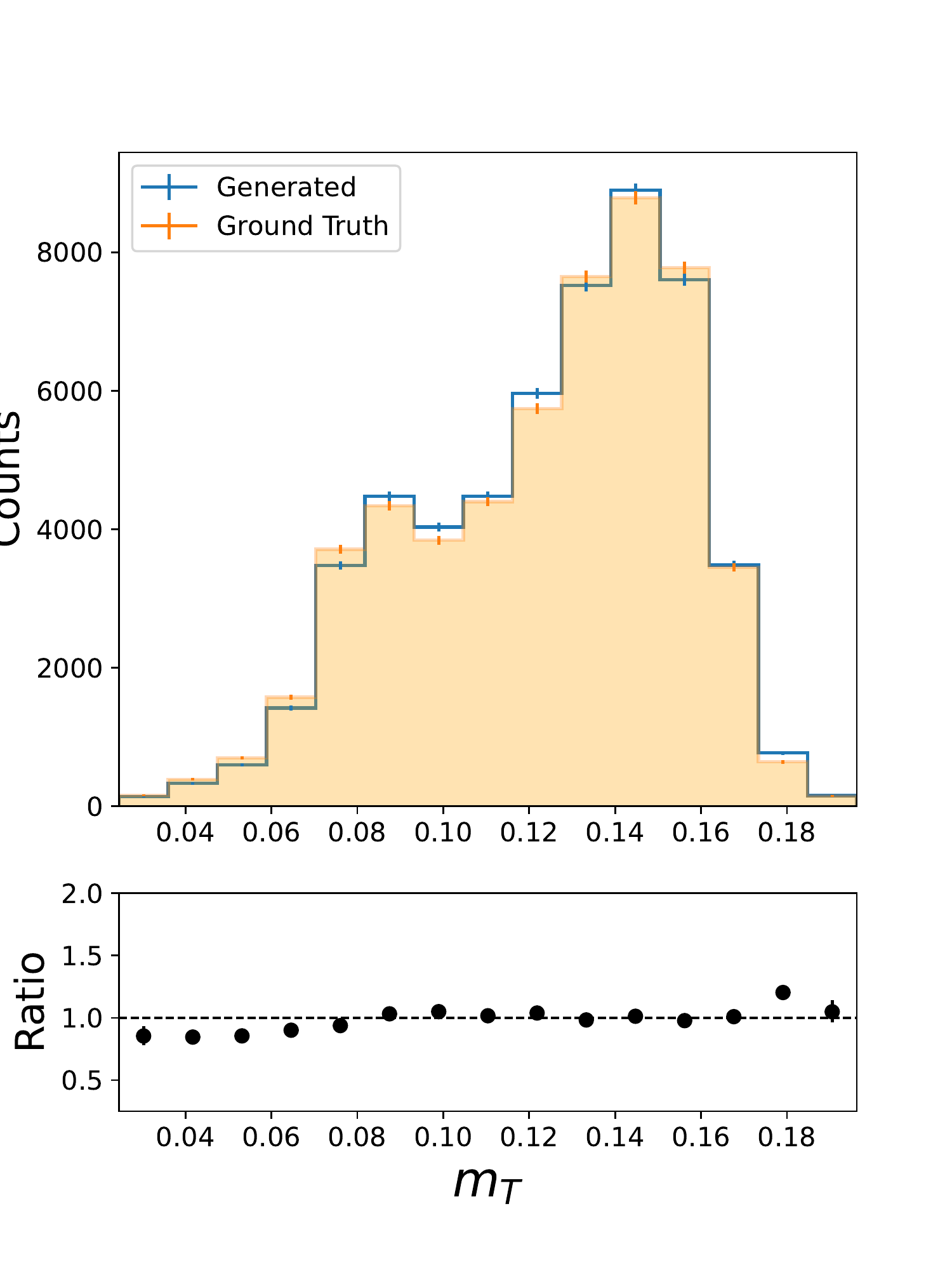}
    
\end{subfigure}

\caption{Mass distribution for the different models on the top-quark dataset. On the left is  VNF, in the middle CNF (m) and to the right CCNF (m). }
\label{fig:mass}
\end{figure}

As mentioned previously, the invariant mass is a powerful variable to test the correlation between the individual jet constituents. It is shown for the different models in Fig.~\ref{fig:mass}. Significant differences can be observed, which are especially evident in the case of the top-quark sample, where there is more structure in the mass distribution. This structure is an artifact from the imperfect anti-$k_T$ clustering. In some cases the b quark in the top decay chain is not clustered to the jet cone and thus we only see a peak of the W boson. While the vanilla NF does not learn this feature, the other two models reproduce the ground truth reasonably well.

\subsection{Performance for variable amount of particles}

\begin{figure}
    \centering

\begin{subfigure}{0.49\textwidth}
    \centering
    \includegraphics[width=\textwidth]{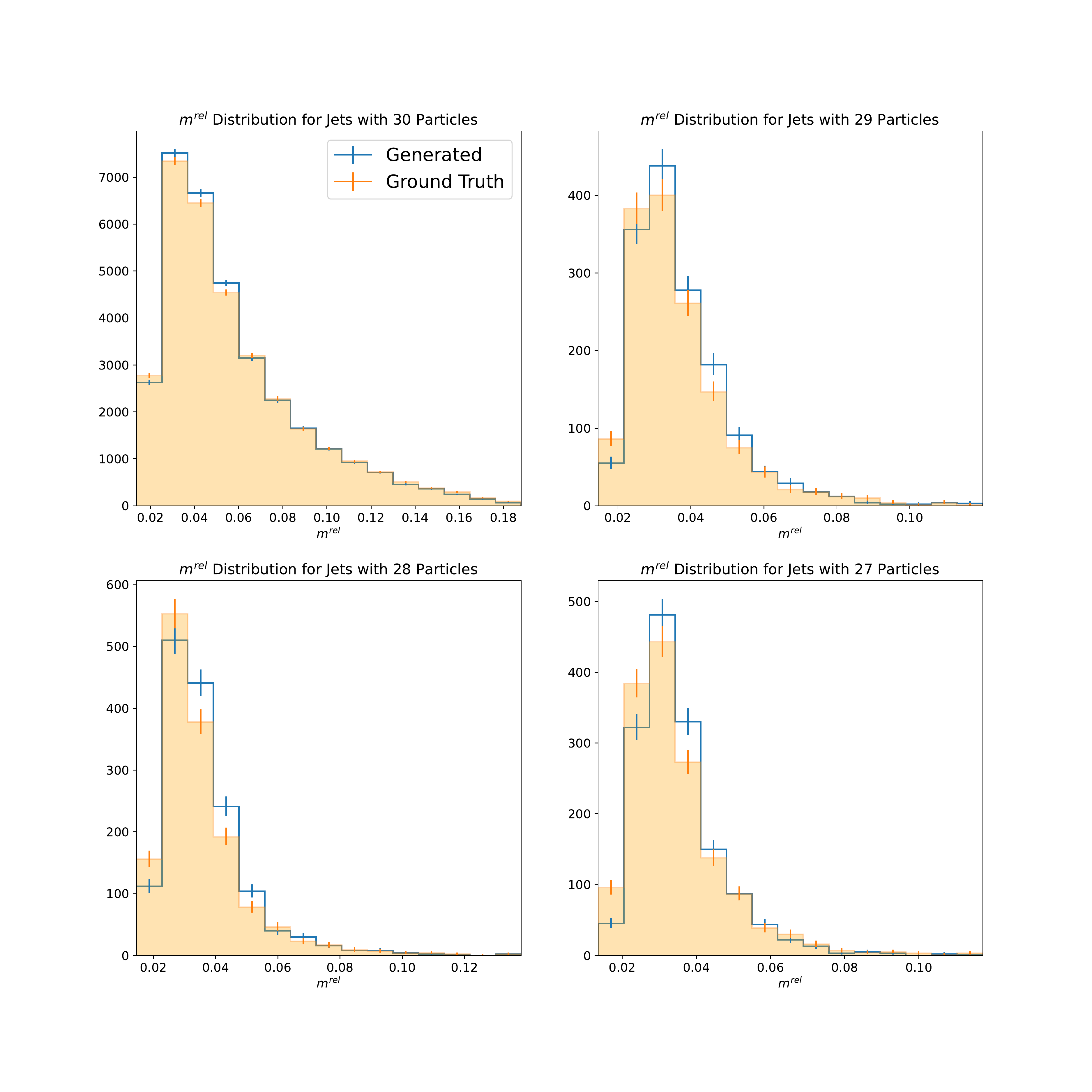}

\end{subfigure}
\begin{subfigure}{0.49\textwidth}
    \centering
    \includegraphics[width=\textwidth]{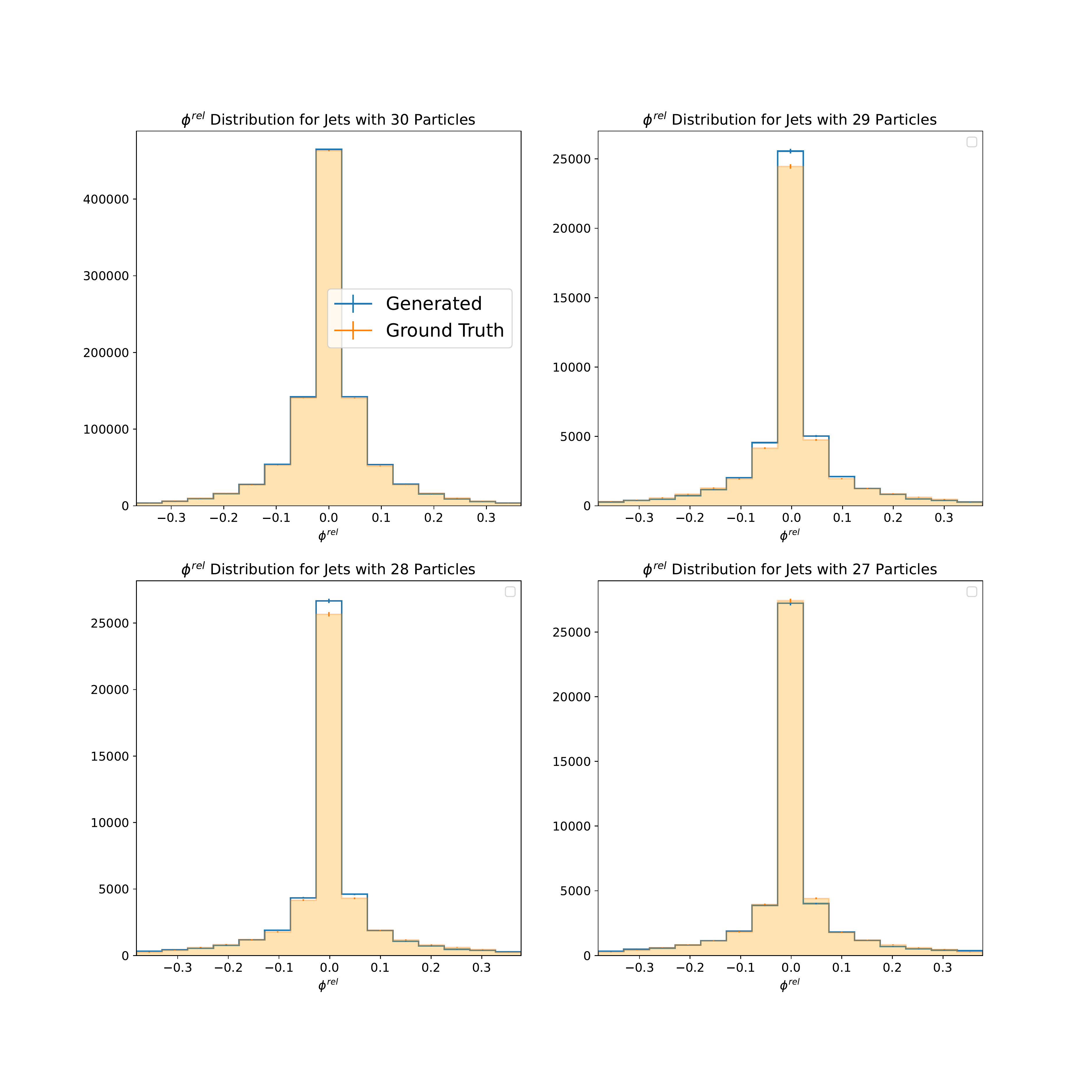}

\end{subfigure}
\caption{Histograms of $m^\mathrm{rel}$ and $\phi^\mathrm{rel}$ for jets with the most frequent number particles for the light-quark dataset for the CNF(m,n) architecture. The plots show that the model is also able to generate jets with fewer than 30 particles. Although there are some deviations, its likely that those are caused by the lack of training data in this regime. }
    \label{fig:variable}
\end{figure}

Figure~\ref{fig:variable} shows samples where the number of particles is restricted to numbers below 30. Results are shown for the light-quark dataset, as we expect the highest amount of variation in the number particles per jet in this case. 

\subsection{Performance while oversampling}

\begin{figure}[h]
    \centering
    \includegraphics[width=\textwidth]{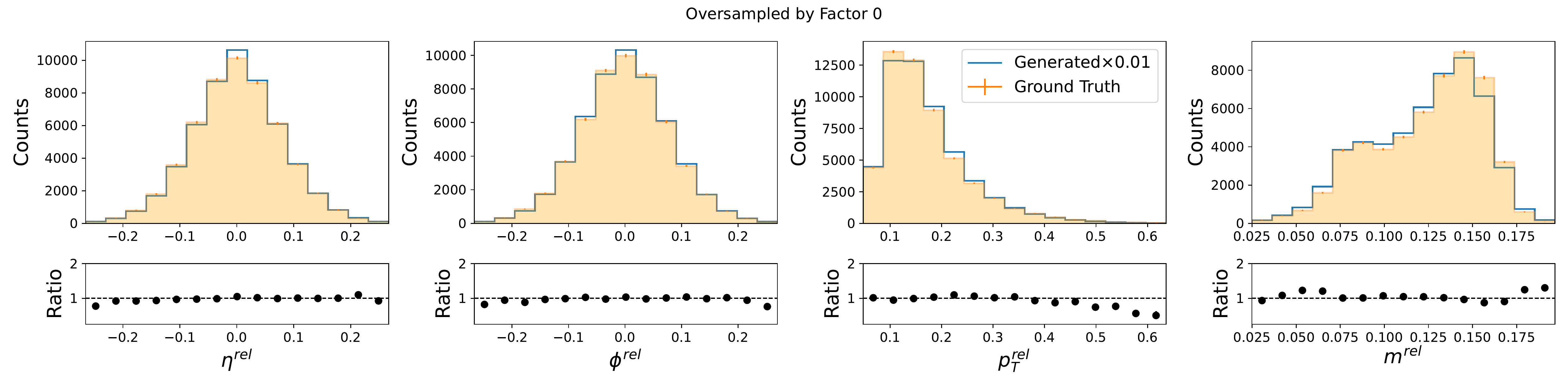}
    \caption{Inclusive distributions of $(\eta^\mathrm{rel},\phi^\mathrm{rel},p_T^\mathrm{rel})$ and the  distribution of calculated mass when 5`338`400 samples were drawn. The generated events are weighted with $\frac{1}{1000}$, as the size of the validation set is 53'308. The model is trained on 106'510 jet samples. }
    \label{fig:oversample}
\end{figure}

One major question in generative modelling is how well they can be over-sampled, meaning how the generated data behaves when more data is sampled than the amount of data that was used for training the model. From a classical statistics viewpoint, it is expected that no more information can be extracted than already present in the training set, as described in~\cite{ganbad}. Other researchers claim that generative models can act as a fit, and thus are able to generate nearly unlimited amount of new data, as discussed in~\cite{gregorfits}. Thus in Fig.~\ref{fig:oversample} the mass and inclusive distributions are compared for the case where about 100 times more samples are drawn than present in the training dataset. No severe deviations can be observed in this case.

\subsection{Correlation}
\begin{figure}[!h]
    \centering

\begin{subfigure}{0.8\textwidth}
    \centering
    \includegraphics[width=\textwidth]{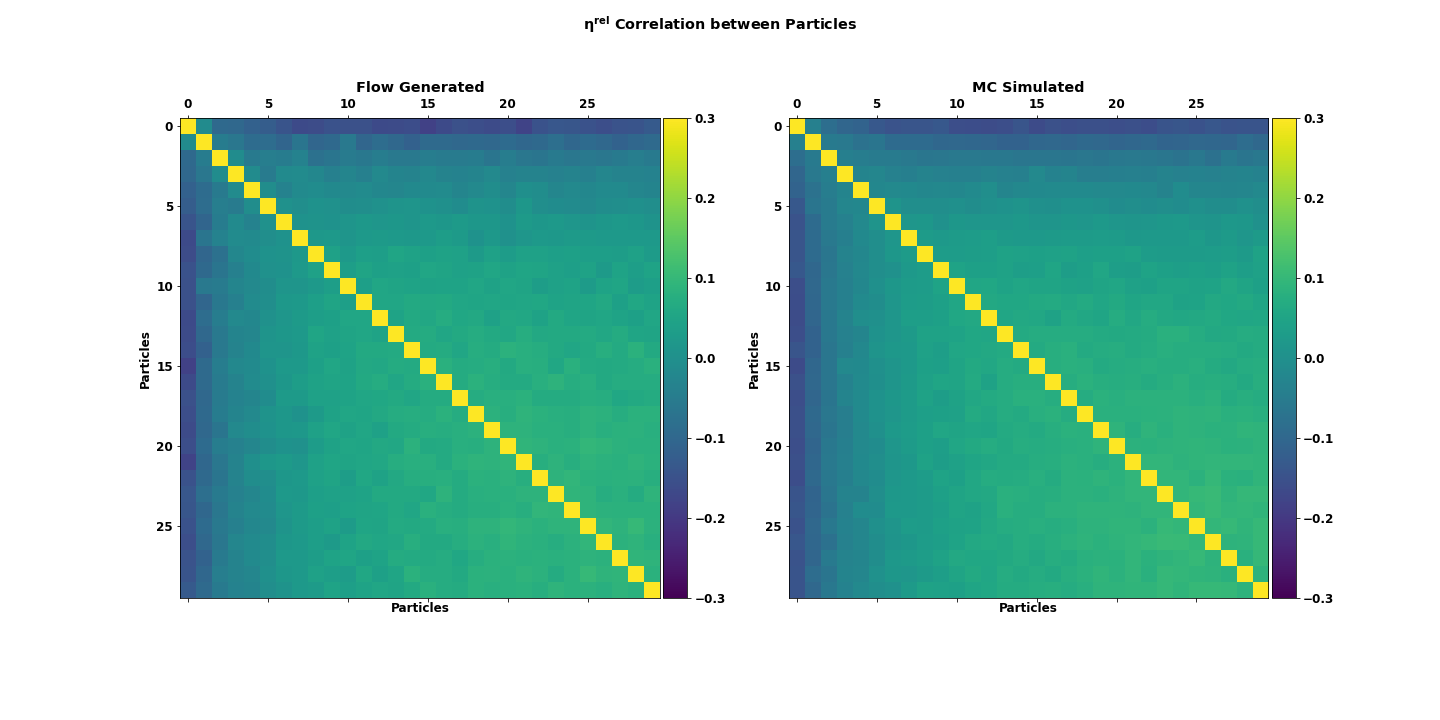}

\end{subfigure}
\begin{subfigure}{0.8\textwidth}
    \centering
    \includegraphics[width=\textwidth]{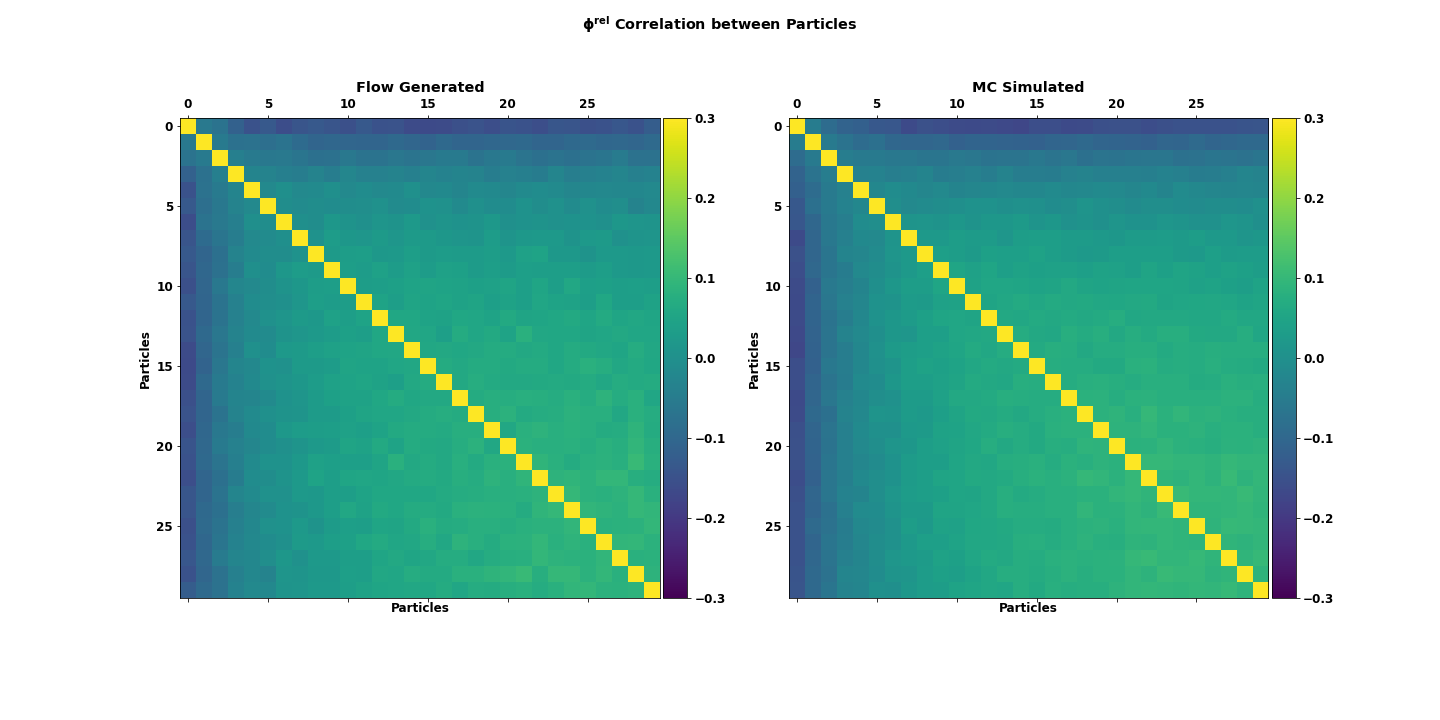}

\end{subfigure}
\begin{subfigure}{0.8\textwidth}
    \centering
    \includegraphics[width=\textwidth]{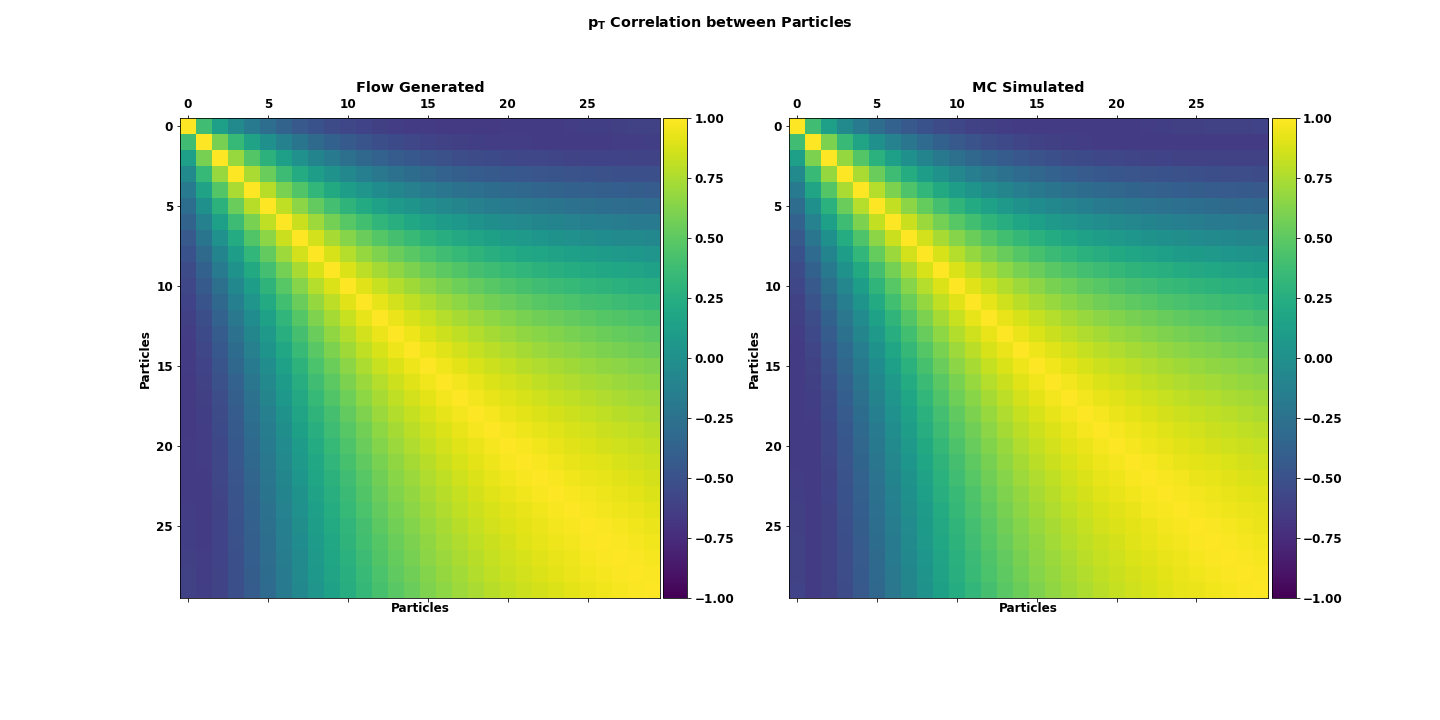}

\end{subfigure}
\caption{Heatmap of linear correlations between particle features. The correlations are shown for all 30 particles and all 3 features. No significant differences can be observed between the MC-generated and flow-generated jets on the top-quark dataset. The CCNF(m,n) model was used.}
    \label{fig:corr}
\end{figure}

Another sensitive control that can be done to compare the ground truth and the generated data is to compare the (linear) correlations between the features of all particles. These are shown in figure~\ref{fig:corr} for the different cases for the top-quark dataset.

\section{Discussion}
\label{sec:discussion}
The results from the previous section show that Vanilla NF, whilst giving acceptable results on the individual features and inclusive distributions, are unable to model more complicated correlations between the particles. This can be due to the construction of the coupling layers, as only 50 \% of all features are seen by the networks in the coupling layers, thus making it more difficult for the network to pick up the correlations. Supplying the mass as a condition and constraining certainly helps but it raises some questions regarding the importance of the latent representation. One could expect that different mass values are encoded at different regions in the latent space. However, if the mass constraint is used the latent representation must be indifferent to the mass. 
 The results from the model are comparable to the state of the art, but when compared to not problem-specific GANs it is evident that the NF models outperform the GANs. Additionally, the training of the NF is trivial compared to the training of GANs as the negative log-likelihood loss yields a stable convergence behavior.
\section{Conclusion}
\label{sec:conclusion}
In this study 3 different architectures for the generation of variable-sized jets are presented. Whilst the most basic rational quadratic spline flows are unable to model the correlations present in the training data, conditioning and constraining the flow on the invariant mass of the jet gives results comparable to the state of the art. As the conditions need to be supplied during sampling, an additional simple model for sampling a two-dimensional distribution is presented as well. The model is evaluated on 6 different metrics, and various control plots between the "real" Monte Carlo simulated data and the flow-generated data are shown. 

\section*{Acknowledgements}

Benno K\"ach is funded by Helmholtz Association’s Initiative and Networking Fund through Helmholtz AI (grant number: ZT-I-PF-5-64).

Moritz Scham is funded by Helmholtz Association’s Initiative and Networking Fund through Helmholtz AI (grant number: ZT-I-PF-5-3).

Alexi Verney-Provatas is partly funded by the European Union’s Horizon 2020 Research and Innovation Programme under the Marie Sklodowska-Curie COFUND scheme with grant agreement No.~101034267.

This research was supported in part through the Maxwell computational resources operated at Deutsches Elektronen-Synchrotron DESY (Hamburg, Germany).
The authors acknowledge support from Deutsches Elektronen-Synchrotron DESY (Hamburg, Germany), a member of the Helmholtz Association HGF.

\newpage
\bibliographystyle{unsrtnat}
\bibliography{references}  






\end{document}